%% file: ms.tex
\documentclass[12pt,preprint]{aastex}

\usepackage{natbib}
  \renewenvironment{thebibliography}[1]{%
    \begin{oldthebibliography}{#1}%
      \setlength{\parskip}{0ex}%
      \setlength{\itemsep}{0ex}%
  }%
  {%
    \end{oldthebibliography}%
  }

\newcommand{\kms}{km s$^{-1}$}
\newcommand{\kmsmpc}{{\rm km\ s\ }^{-1}{\rm\ Mpc}^{-1}}

\newcommand{\gtrsi}{\mathrel{\hbox{\rlap{\hbox{\lower4pt\hbox{$\sim$}}}\hbox{$>$}}}}
\newcommand{\lesssi}{\mathrel{\hbox{\rlap{\hbox{\lower4pt\hbox{$\sim$}}}\hbox{$<$}}}}

\slugcomment{Received: 2007 July 21;  Accepted:  2007 August 10}

\shorttitle{Constraining the Type Ia Supernova Progenitor} \shortauthors{Leonard}

\received{2007 July 21}
\begin{document}

\slugcomment{Submitted: 2007 July 21;  Accepted:  2007 August 10}

\title{Constraining the Type Ia Supernova Progenitor:  The Search for Hydrogen
  in Nebular Spectra\footnote{Some of the data presented herein were obtained
  at the W.M. Keck Observatory, which is operated as a scientific partnership
  among the California Institute of Technology, the University of California
  and the National Aeronautics and Space Administration. The Observatory was
  made possible by the generous financial support of the W.M. Keck Foundation.
  Additional observations were obtained at the Gemini Observatory, which is
  operated by the Association of Universities for Research in Astronomy, Inc.,
  under a cooperative agreement with the NSF on behalf of the Gemini
  partnership: the National Science Foundation (United States), the Particle
  Physics and Astronomy Research Council (United Kingdom), the National
  Research Council (Canada), CONICYT (Chile), the Australian Research Council
  (Australia), CNPq (Brazil) and CONICET (Argentina).}}

\vspace{2cm}

\author{Douglas C. Leonard}
\affil{Department of Astronomy, San Diego State University, San Diego, CA 92182-1221}
\email{leonard@sciences.sdsu.edu}

\vspace{1cm}

\begin{abstract}

Despite intense scrutiny, the progenitor system(s) that gives rise to Type Ia
supernovae remains unknown.  The favored theory invokes a carbon-oxygen white
dwarf accreting hydrogen-rich material from a close companion until a
thermonuclear runaway ensues that incinerates the white dwarf.  However,
simulations resulting from this single-degenerate, binary channel demand the
presence of low-velocity H$\alpha$ emission in spectra taken during the late
nebular phase, since a portion of the companion's envelope becomes entrained in
the ejecta.  This hydrogen has never been detected, but has only rarely been
sought. Here we present results from a campaign to obtain deep, nebular-phase
spectroscopy of nearby Type Ia supernovae, and include multi-epoch observations
of two events: SN~2005am (slightly subluminous) and SN~2005cf (normally
bright).  No H$\alpha$ emission is detected in the spectra of either object.
An upper limit of $0.01\ M_\odot$ of solar abundance material in the ejecta is
established from the models of \citet{Mattila05} which, when coupled with the
mass-stripping simulations of \citet{Marietta00} and \citet{Meng07},
effectively rules out progenitor systems for these supernovae with secondaries
close enough to the white dwarf to be experiencing Roche lobe overflow at the
time of explosion.  Alternative explanations for the absence of H$\alpha$
emission, along with suggestions for future investigations necessary to
confidently exclude them as possibilities, are critically evaluated.

\end{abstract}

\medskip

\section{Introduction}
\label{sec:1}

Ever since the currently favored single-degenerate, binary channel was proposed
as the progenitor system for Type Ia supernovae \citep{Whelan73}, all models of
the impact of the exploded white dwarf (WD) on the secondary star have
indicated that significant amounts of solar-abundance material, stripped from
the secondary's envelope, become entrained in the ejecta
\citep{Wheeler75,Fryxell81,Taam84,Chugai86,Livne92,Marietta00,Meng07}.
Observational evidence for this material, however, still eludes us
\citep{Mattila05}, and serves as one reminder among many that we still lack
direct observational proof for the single-degenerate scenario
\citep{Branch95,Livio01}.

The most detailed theoretical investigation of the expected amount and
distribution of stripped material within a young Type Ia supernova (SN~Ia)
remnant is that of \citet{Marietta00}, who studied the problem with
two-dimensional numerical simulations.  Four basic progenitor systems were
investigated, including three with secondaries (a main-sequence, subgiant, or
red giant star) close enough for mass-transfer to occur through Roche lobe
overflow (RLOF), and one containing a secondary (a red giant) donating material
through a strong stellar wind --- the symbiotic case.  All secondaries were
given masses of $\sim 1\ M_\odot$ at the time of the explosion and were placed
either just within (the RLOF cases), or just beyond (the symbiotic case), the
limiting distance from the WD within which RLOF can occur (i.e., $a/R = 3$,
where $a$ is the orbital separation in units of the secondary star's radius,
$R$; see \citealt{Eggleton83}).  Three additional systems, in which a
main-sequence secondary was placed too far away to experience RLOF (thus
rendering it an unlikely progenitor system) were also included to establish the
scaling between orbital separation and amount of stripped material.

The numerical results of the \citet{Marietta00} study confirmed predictions
from earlier analytic work \citep[e.g.,][]{Wheeler75,Chugai86} that substantial
material is indeed stripped, with the amount ranging from a minimum of $0.15\
M_\odot$ for a close ($a/R = 3$) main-sequence secondary, up to nearly the
entire envelope ($\sim 0.5\ M_\odot$) for a similarly placed red giant.
Increasing the orbital separation beyond the RLOF limit in the
\citet{Marietta00} simulations resulted in a dramatic decrease in the amount of
stripped material: For $a/R = 12$, a main-sequence secondary loses only
$0.0018\ M_\odot$.  However, since such systems lack an efficient mechanism for
mass transfer, they are not considered to be viable SN~Ia progenitors.  A red
giant secondary placed at a similarly distant location, however, could
potentially donate mass through a strong stellar wind, making the symbiotic
case a possibility if large orbital separations are required
\citep[e.g.,][]{Munari92}.

As discussed by \citet{Meng07}, one shortcoming of the \citet{Marietta00} study
is their use of standard solar-model stars for the companion, rather than
companions whose structures have been appropriately modified due to having
evolved in a binary system \citep[e.g.,][]{Eggleton73}.  To investigate the
effect this simplification might have had on the results, \citet{Meng07} use an
analytic model to estimate the amount of mass expected to be stripped from a
variety of evolved secondaries.  (Their analytic approach was first tested
using the unevolved secondaries used by \citealt{Marietta00}, and was
demonstrated to approximate the results obtained numerically.)  The result is
that the quantity of material expected to be stripped from evolved secondaries
is considerably lower than that predicted for standard solar-model companions.
In fact, \citet{Meng07} find that the minimum value for systems experiencing
RLOF is diminished from $0.15\ M_\odot$ to only $0.035\ M_\odot$.  The
reduction arises primarily from the pre-explosion mass-loss producing a more
compact companion star whose material is more difficult to strip than it is in
the unevolved case.  \citet{Meng07} stress, however, that their new values are
really lower bounds on the amount of stripped material, since their analytic
approach does not consider the thermal energy imparted by the ejecta to the
companion envelope, which likely serves to heat and vaporize a portion of it
and thereby increase the amount of stripped material
\citep[e.g.,][]{Fryxell81,Mattila05}.  Thus, $0.035\ M_\odot$ serves as a
conservative lower bound on the expected amount of stripped material resulting
from their models.

The typical velocity of the stripped material is found in all studies to be far
slower than the $\sim 10,000$~\kms\ velocity that characterizes the bulk of the
iron-rich ejecta \citep{Chugai86,Marietta00,Meng07}.  This has the effect of
placing it almost entirely in the central region of the supernova remnant, with
the majority of it predicted to be moving with a velocity of under
$1,000$~\kms\ \citep{Marietta00}.  The stripped material is largely confined to
the downstream region behind the companion star, where it contaminates a solid
angle that ranges from $66^\circ$ for the main-sequence companion to
$115^\circ$ for the red-giant companion \citep{Marietta00}.

The low velocity of the stripped material renders it undetectable when the
faster-moving, iron-rich ejecta are optically thick.  However, detailed
radiative transfer calculations performed by \cite{Mattila05} predict that it
should become visible at late times, when the outer ejecta have thinned out and
become transparent enough to reveal the slower-moving gas in the central
regions.  The most prominent expected spectral signature of the companion
star's stripped material is narrow H$\alpha$ emission in nebular spectra
\citep{Mattila05}, taken more than $\sim 250$ days after maximum light.  The
H$\alpha$ emission should be present within $\pm 1,000$~\kms\ {\rm of\ }
$\lambda_0 = 6563$ \AA\ but, due to the expected asymmetry in the distribution
of the solar-abundance material, could present an H$\alpha$ profile ranging
from a very narrow spike to a broader emission line in the observed spectrum.
Obtaining spectra with high enough resolution to resolve fairly narrow lines is
thus a useful component of a targeted search for this H$\alpha$.

To date, only a few nebular SN~Ia spectra have been obtained and H$\alpha$ has
never been detected, although the majority of the spectra lack the spectral
resolution and signal-to-noise ratio to place interesting constraints on
the companion.  The best limits, by far, come from the recent study by
\cite{Mattila05}, which constrains hydrogen-rich material in SN~2001el to be
$\lesssi 0.03\ M_\odot$ from a low-resolution ($\sim 700$~\kms) spectrum
obtained 398 days past maximum light.

In an effort to expand and improve on earlier work, both in terms of the
number of objects studied as well as the resolution, sensitivity, and temporal
coverage of the spectra, we have initiated a program to obtain deep, moderate
resolution ($\lesssim 150$~\kms, or $\sim 3$~\AA, at H$\alpha$), late-time
spectra of SNe~Ia at multiple epochs using the Keck and Gemini telescopes.  The
first phase of this project has garnered data on two objects: SN~2005am, a
slightly subluminous \citep{Li06} event and SN~2005cf, an SN~Ia of normal
brightness \citep{Pastorello07}.  We present and analyze our observations in
\S~\ref{sec:2}, discuss the results in \S~\ref{sec:3}, and conclude in
\S~\ref{sec:4}.  

\section{Observations and Analysis}
\label{sec:2}

We obtained a total of five deep nebular-phase spectra: Two for SN~2005am and
three for SN~2005cf.  Details of the observations are given in
Table~\ref{tab:1}.  Following initial processing of the frames,\footnote{For
Gemini observations, the frames were processed using the tasks {\it gprepare},
{\it gsreduce}, {\it gsflat}, and {\it gmosaic} in the Gemini IRAF package.} we
extracted all one-dimensional sky-subtracted spectra optimally \citep{Horne86}
in the usual manner using the {\it apall} task within IRAF.\footnote{IRAF is
distributed by the National Optical Astronomy Observatories, which are operated
by the Association of Universities for Research in Astronomy, Inc., under
cooperative agreement with the National Science Foundation.}  Each spectrum was
then wavelength and flux calibrated, and corrected for continuum atmospheric
extinction and telluric absorption bands \citep[][and references
therein]{Matheson00}.  Careful examination of the two-dimensional spectra
reveals only faint night-sky emission at the nominal location of H$\alpha$, and
so we deem the potential for contamination at these wavelengths due to improper
background (``sky'') removal to be negligible.

\begin{figure}
\scalebox{0.9}{
\plotone{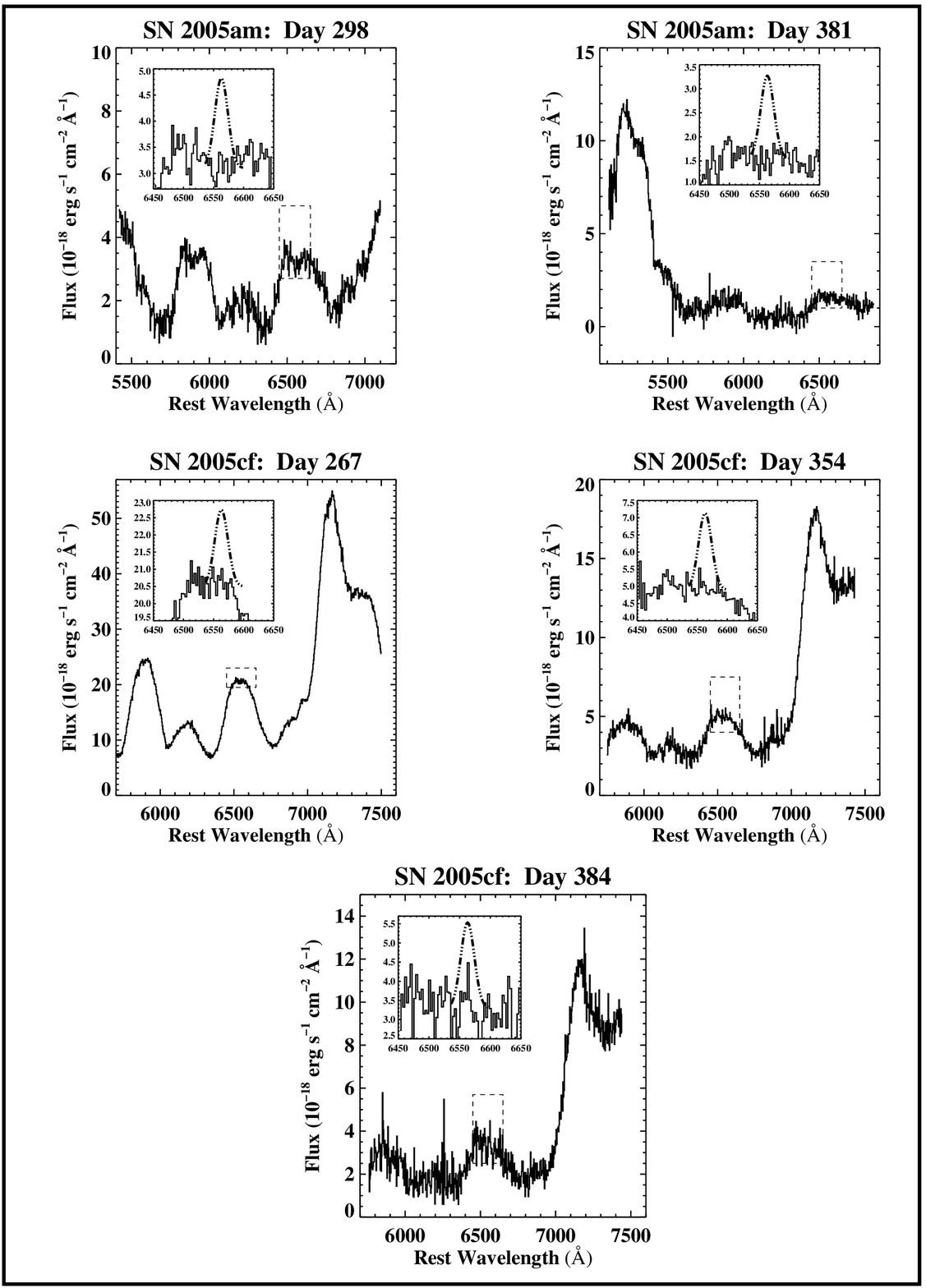} 
              }
\vskip -0.2in
\caption{Late-time spectra of two SNe~Ia, with day since $B$ maximum indicated.
  The spectra are displayed at 3 \AA\ bin$^{-1}$, the approximate resolution at
  H$\alpha$.  The expected strengths of the H$\alpha$ line resulting from
  $0.05\ M_\odot$ of solar-abundance material according to the day 380 models
  of \cite{Mattila05} are shown as dot-dashed lines in the {\it insets}; note
  that since the H$\alpha$ emission is a time-dependent phenomenon, the
  estimated strength of these lines in the day 298 and day 267 spectra of
  SN~2005am and SN~2005cf, respectively, are only approximate. 
\label{fig:1} }
\end{figure}

The final spectra are displayed in Figure~\ref{fig:1}.  To search for H$\alpha$
emission, or to place limits on the amount of solar-abundance material present
in the inner ejecta, we subjected each spectrum to the following analysis
procedure.

First, we attempted to place each spectrum on an absolute flux scale.
   Adjustments to the original flux levels are necessary since most
   observations were made with a rather narrow slit width compared with the
   seeing (see Table~\ref{tab:1}), which makes the flux level susceptible to
   the effects of seeing variations between the SN observations and those of
   the flux standard star; some observations were also not obtained under
   photometric conditions.  To produce an approximate absolute flux
   calibration, then, we computed synthetic photometry on our spectra and
   compared it with estimates of the $V-$ or $R-$band (depending on the
   spectral range of the spectrum) magnitudes of the SNe at the spectral
   epochs.  Since none of our spectra cover an entire passband, we extended
   each spectrum's range by joining it with SN~Ia spectra of similar age from
   the SUSPECT\footnote{see
   http://bruford.nhn.ou.edu/$\sim$suspect/index1.html} database,\footnote{The
   specific spectra used for this purpose include SN~1998bu (day 329;
   \citealt{Cappellaro01}), SN~2003cg (day 385; \citealt{Elias-Rosa06}), and
   SN~2002bo (day 375; \citealt{Benetti04}).}  scaled to match the flux of our
   spectra in regions of overlap.  We then compared our synthetic photometry
   with estimates of the actual brightnesses of the SNe.  Since no published
   late-time (i.e., beyond $t > 100$ days) photometry exists for either
   SN~2005am or SN~2005cf, but high-quality early-time photometry does
   \citep{Li06, Pastorello07}, we derived approximate $V$ and $R$ magnitudes of
   the SNe at our spectral epochs by first extrapolating the early-time light
   curves to day 200 through comparison with the light curves of SN~2003du
   \citep{Stanishev07}, and then applying the late-time Type Ia decay-rates
   reported by \citeauthor{Lair06} (\citeyear{Lair06}; $\Delta V = 1.46 {\rm\
   mag\ 100\ day}^{-1},\ \Delta R = 1.54 {\rm\ mag\ 100\ day}^{-1}$).  We then
   adjusted each of our spectra by multiplying it by the scale factor needed to
   bring its synthetic magnitude to the estimated actual apparent magnitude.
   The only spectrum for which we did not apply this procedure was the day 298
   epoch of SN~2005am, since the night was photometric, the flux of the SN from
   one exposure to the next was very consistent (indicating non-varying seeing
   throughout the observation sequence), and the spectrum of a nearby star
   (HD~79289) taken immediately after the SN observations was found to have a
   synthetic $V$ magnitude within $16\%$ of the value reported by
   SIMBAD;\footnote{see http://simbad.u-strasbg.fr/simbad/} the scale factor
   for this epoch was thus taken to be the value needed to place HD~79289 on a
   correct absolute scale.  The scale factors applied to all spectra are given
   in Table~\ref{tab:2}.  Given the uncertainties inherent in our technique,
   the absolute flux calibrations are probably accurate to $\sim 25\%$, in
   general.

We next removed the redshift and rebinned each rest-frame spectrum to 3 \AA
${\rm\ bin}^{-1}$, the approximate spectral resolution (see Table~\ref{tab:1}).
For SN~2005am in NGC~2811, we adopted the NASA/IPAC Extragalactic Database
(NED) recession velocity of $2,368$~\kms, and for SN~2005cf in MCG-01-39-3 we
adopted the NED recession velocity of $1,937$~\kms.

We then searched for narrow H$\alpha$ emission within $\pm 1000$~\kms\ ($\pm
22$ \AA) of $\lambda_0 = 6563$~\AA\ in each rescaled, rest-frame spectrum by
smoothing the spectrum with a second-order Savitsky-Golay smoothing polynomial
\citep{Press92} of width $\sim 100$ \AA, differencing the smoothed and
unsmoothed spectra, and then examining the residuals for narrow emission near
H$\alpha$.  This procedure yielded no evidence for narrow H$\alpha$ emission in
any spectrum.  Indeed, no unexpected narrow features at any wavelength
(including spectral regions near [\ion{O}{1}] $\lambda\lambda 6300, 6364$ and
[\ion{Ca}{2}] $\lambda\lambda 7291, 7324$, which might be expected to produce
features similar to H$\alpha$ at these epochs from stripped solar-abundance
material) were found.

To determine the greatest strength (i.e., equivalent width) of a feature that
could have remained undetected at the location of H$\alpha$ in each spectrum,
we adopted the procedure of \citet{Leonard4}, who report the $3 \sigma$ lower
bound on the equivalent width of an undetected feature in a spectrum to be:

\begin{equation}
W_\lambda(3\sigma) = 3 \Delta\lambda\ \Delta I \sqrt{W_{line} / \Delta\lambda\
} \sqrt{1 / B},
\label{eqn:1}
\end{equation}

\noindent where $\Delta \lambda$ is the width of a resolution element (in \AA),
$\Delta I$ is the 1$\sigma$ root-mean-square fluctuation of the flux around a
normalized continuum level, $W_{line}$ is the full-width at half-maximum of the
expected line feature (in \AA) and $B$ is the number of bins per resolution
element in the spectrum.  The values for $\Delta \lambda$, $\Delta I$, and $B$,
along with the computed values of $W_\lambda(3\sigma)$, are reported in
Table~\ref{tab:2}.  The width of the line feature, $W_{line}$, was assumed to be
$22$ \AA.

While many studies have predicted the existence of narrow H$\alpha$ in nebular
spectra of SNe~Ia, only \citet{Mattila05} gives a quantitative estimate of its
expected strength.  Thus, to translate our detection thresholds into estimates
of the maximum amount of solar-abundance material that could have given rise to
an H$\alpha$ line too weak to have been detected, we rely on this study alone.

In the \citet{Mattila05} study, the code described by \citet{Kozma05} is
modified to artificially include varying amounts of solar abundance material in
the inner ($\pm 1,000$~\kms) region of the ejecta.  The resulting spectra were
computed by \citet{Mattila05} in a time-dependent, one-dimensional, spherically
symmetric\footnote{Note that while the expected asymmetry in the location of
the stripped material would undoubtedly alter the shape of the resulting line
profile, \citet{Mattila05} conclude that it should not affect the total
strength of the line, and so the assumption of spherical symmetry is deemed to
be acceptable for the purposes of their study (and, hence, ours).} manner for
an SN~Ia 380 days after maximum light.  From the parameters given by
\citeauthor{Mattila05} (see their Figure~6 and associated discussion), we
estimate that $0.05\ M_\odot$ of solar-abundance material produces a
luminosity in the peak of the H$\alpha$ line of $\sim 3.36 \times 10^{35} {\rm\
erg\ s}^{-1} {\rm\ \AA}^{-1} $.

Translating predicted H$\alpha$ luminosity into observed H$\alpha$ strength
requires estimates of the distance and extinction of the SNe.  For SN~2005am,
we adopt the NED Hubble flow ($H_0 = 73\ \kmsmpc$) distance of $36.8 \pm 2.6$
Mpc based on the 3k CMB rest frame as determined by \citet{Fixsen96}, and the
total extinction of $A_V = 0.27 \pm 0.08$ mag determined by \citet{Li06}.  For
SN~2005cf, we use the distance and extinction values adopted by
\citet{Pastorello07} of $D = 31.77 \pm 4.8 {\rm\ Mpc}$ and $A_V = 0.32 \pm
0.03$ mag.  For $0.05\ M_\odot$ of solar-abundance material 380 days after
explosion, we derive expected H$\alpha$ line fluxes at the profile's peak of
$1.72 \times 10^{-18} $ and $2.23 \times 10^{-18} {\rm\ erg\ s}^{-1} {\rm\
\AA}^{-1} {\rm\ cm}^{-2}$ for SN~2005am and SN~2005cf, respectively.

To derive the equivalent width that such features would have had in our
observed spectra, we approximated the H$\alpha$ emission as a Gaussian with a
$1\sigma$ width of $11$~\AA, and placed the resulting profile on top of the
estimated ``continuum'' level at $\lambda = 6563$ \AA\ (see Fig.~\ref{fig:1}).
We then measured the equivalent width of these theoretical profiles within the
region $\pm 11$~\AA\ from $\lambda_0 = 6563$ \AA\ [the same spectral range used
to derive $W_\lambda(3\sigma)$ earlier] in each spectrum.  The calculated
values for $W_\lambda (0.05 M_\odot)$ are given in Table~\ref{tab:2}.  Since
the H$\alpha$ emission is clearly a time-dependent phenomenon and the
\citet{Mattila05} model is specifically for day~380, our estimate of the
strength of the line in the day 298 and day 267 spectra of SN~2005am and
SN~2005cf, respectively, are quite crude.  Fortunately, since the H$\alpha$
emission is powered primarily by gamma-ray deposition \citep{Mattila05}, and
the optical depth to gamma-rays in the central, hydrogen rich region should be
higher at earlier times, we would expect the H$\alpha$ line to be even stronger
at earlier times than it is on day 380 (assuming, of course, that the outer,
iron-rich ejecta are transparent at earlier times, and allow us to see it; see
\S~\ref{sec:3}).  Thus, our calculated value of $W_\lambda (0.05\ M_\odot)$ for
epochs earlier than 380 days should serve as a conservative lower limit for
these spectra.

Finally, we assume a linear scaling between the amount of stripped material and
the expected equivalent width of the H$\alpha$ emission line (an approximation
justified through examination of Figure~6 of \citealt{Mattila05}, where
theoretical profiles are shown for $0.01$ and $0.05\ M_\odot$) to
arrive at our final estimate for the upper limit on the amount of solar
abundance material that could have remained undetected at each spectral epoch.
We calculated this limit according to:

\begin{equation}
{\rm M} (M_\odot) \leq \frac{W_\lambda(3\sigma)}{W_\lambda (0.05 M_\odot)}
\times 0.05 .
\label{eqn:2}
\end{equation}

Final results are listed in Table~\ref{tab:2}, and can be summarized
succinctly: We restrict the amount of solar-abundance material that could have
evaded detected to be $< 0.01\ M_\odot$ at all spectral epochs except day 384
of SN~2005cf, for which the formal upper limit is $0.02\ M_\odot$.

\section{Discussion}
\label{sec:3}

When coupled with the modeling results of \citet{Mattila05} and the
mass-stripping estimates of \citet{Marietta00} and \citet{Meng07}, our
restrictions on H$\alpha$ emission place strong constraints on the progenitor
systems that could have given rise to SN~2005am and SN~2005cf.  In short, they
rule out all hydrogen-donating companions close enough to the WD to
have been experiencing RLOF at the time of explosion (i.e., $a/R
< 3$; see \S~\ref{sec:1}).  Under the favored single-degenerate scenario, this
leaves only widely separated systems in which the companion (i.e., a red giant)
donates matter through a strong stellar wind, as a viable option.  For the
symbiotic case, if we assume that the same scaling between the amount of
stripped material and the orbital separation holds for red giants as was found
by \cite{Marietta00} for main-sequence secondaries (see \S~\ref{sec:1}; note
that red giant companions were not analyzed by \citealt{Meng07}), then we
conclude that the mass-donating red giant must be $\gtrsim 10$ A.U. from the
WD to be consistent with our H$\alpha$ nondetections.  For a $1.4\
M_\odot$ primary and a $1.0\ M_\odot$ secondary, this corresponds to an orbital
period of $\gtrsim 20$ yr.

Our main conclusion --- that the most likely progenitors of SN~2005am and
SN~2005cf were widely separated symbiotics --- stands in evident contrast to
the work of \citet{Panagia06} on the search for, and subsequent nondetection
of, radio emission from SNe~Ia.  In their study, \citet{Panagia06} set upper
limits on mass-loss rates of $\sim 10^{-7}\ M_\odot\ {\rm yr}^{-1}$ for the
progenitor systems.  Since this is of the same order of magnitude as the
observed mass-loss rates from symbiotics \citep[e.g.,][]{Jung04,Crowley06},
\citet{Panagia06} specifically rule out symbiotics as potential progenitors of
SNe~Ia.  However, the inferences drawn by \citet{Panagia06} rely on the
assumption that SN~Ia radio light curves behave identically to those of
SNe~Ib/c.  Such an assumption is necessary because the correlation between wind
density and radio luminosity for SNe~Ia cannot be calculated from theory at
this point.  In the absence of theoretical or observational support for such an
assumption the very low mass-loss rates claimed by \citet{Panagia06} cannot yet
be considered definitive \citep{Hughes07}.  The radio non-detections may
therefore still permit the symbiotic scenario.  Along these same lines, the
investigation of X-ray non-detections of SNe~Ia by \citet{Hughes07} provides
mass-loss limits of only $\sim 10^{-5}\ M_\odot\ {\rm yr}^{-1}$, which are not
stringent enough to rule out symbiotics.  While a low-density circumstellar
environment is clearly favored by both the radio and X-ray non-detections of
SNe~Ia, it seems premature to rule out progenitor classes on this basis at the
present time.

There are additional caveats to our conclusion that must be given.  Of primary
significance is the fact that its robustness depends critically on the
validity of the theoretical modeling of just a few studies. Taken together,
these studies predict that a substantial amount of stripped hydrogen should
exist at low velocities \citep{Marietta00,Meng07} and emit a strong H$\alpha$
feature at late times \citep{Mattila05}.  There are, however, alternative
explanations for the observed lack of hydrogen that must be thoroughly
investigated to build confidence in the inferences that can be drawn from our
study.  We now critically evaluate four such alternatives, and point out areas
ripe for additional theoretical study.

(1) {\it The hydrogen is ``hidden'' behind an opaque screen of faster-moving
  ejecta.}  Posed simply: Are the fast-moving, iron-rich ejecta truly optically
  thin at late times?  Since permitted \ion{Fe}{2} lines dominate the
  underlying spectrum of SNe~Ia near $6563$ \AA, significant opacity in these
  lines could provide an ``iron curtain'' that shields the hydrogen emission
  from our view.  Indeed, \citet{Kozma05} state that several Fe II lines do
  remain optically thick in their models even at late times.  On the other
  hand, the simulations of \citet{Mattila05}, which predict the strong
  H$\alpha$ feature, are explicitly {\it based} on the models of
  \citet{Kozma05}, and so it seems reasonable to infer that the specific
  \ion{Fe}{2} lines near H$\alpha$ do not provide significant opacity in the
  \citet{Kozma05} models.  Independent confirmation of the transparency of the
  outer ejecta of SNe~Ia at late times is needed.

(2) {\it There is far less hydrogen entrained in the ejecta than current models
    predict.}  By predicting up to a factor of four reduction in the amount of
    stripped mass compared with the models of \citet{Marietta00},
    \citet{Meng07} demonstrate the potentially large effect that binary
    evolution can have on the amount of stripped material.  Might other aspects
    of binary evolution be at work to further reduce the stripped mass?  A
    possibly important mechanism not considered by \citet{Meng07} is the
    outcome that a strong ``accretion wind'', blown from the WD during
    the mass-accretion phase, could have on the secondary's envelope.
    Accretion winds were originally conceived by \citet{Hachisu96} as a means
    to stabilize the mass-transfer process and allow the system to avoid a
    common envelope phase.  Further investigation by \citet{Hachisu99} and
    \citet{Hachisu03} revealed that the winds should also heat and ablate a
    significant amount of material from a close secondary's envelope.  From
    these considerations, \citet{Hachisu99} estimate that the wind could strip
    mass from such a companion at rates as high as $\sim 10^{-6}\ M_\odot\ {\rm
    yr}^{-1}$ for a period of up to $\sim 10^{6}\ {\rm\ yr}$ prior to the
    explosion.  If such a process actually precedes an SN~Ia explosion, it
    would clearly leave the secondary's envelope in a vastly different state
    from the models considered to date.  In fact, such secondaries could
    potentially lose their entire hydrogen envelope and evolve to become
    helium-donators, which would naturally account for a lack of hydrogen in
    late-time spectra \citep{Branch95}.

    There are both theoretical and observational objections to strong WD winds
    playing a large roll in stripping mass from the secondary star in SN~Ia
    progenitor systems, however.  On the theoretical front, \citet{Livio01}
    questions the high efficiency of the mass stripping proposed by
    \citet{Hachisu99}, and also points out that at high accretion rates, the
    bulk of the WD's wind may be strongly collimated in a direction
    perpendicular to the accretion disk rather than in the direction of the
    companion star.  Such a scenario would result in little stripping.
    Observational doubt on accretion winds playing a major role in pre-SN~Ia
    evolution is cast by \citet{Badenes07}, who find that such an optically
    thick outflow from a WD's surface should excavate a large, low-density
    cavity around the system.  Such a large evacuated region is incompatible
    with the known observational properties of SN~Ia remnants in the Galaxy,
    the Large Magellanic Cloud, and M31.  It thus appears that, while such
    accretion winds may occur in nature \citep[e.g.,][]{Hachisu03}, they do not
    evidently precede a majority of SN~Ia explosions.  Further consideration of
    this and other potential mass-stripping mechanisms is certainly warranted.

(3) {\it The hydrogen gas is insufficiently powered to produce H$\alpha$
    emission.}  \citet{Mattila05} conclude that the hydrogen-rich, central
    region of the young supernova remnant presents high optical depth to
    gamma-ray photons that originate from the radioactive decay of isotopes
    synthesized in the explosion.  The gamma-ray trapping is then responsible
    for powering the H$\alpha$ line emission.  This mechanism is quite
    different from the situation in the outer, lower density regions of the
    ejecta, whose optical emission is predominantly powered by local positron
    deposition \citep[e.g.,][]{Stritzinger07}.  Since the production of a
    strong H$\alpha$ line from even small amounts of solar-abundance material
    relies fundamentally on the inner ejecta's ability to trap gamma rays, the
    high opacity of the material, and the existence of sufficient numbers of
    gamma-rays to power the emission, should be confirmed by independent
    modeling.

(4) {\it The single-degenerate scenario is not responsible for these SNe~Ia.}
    A double-degenerate progenitor system, in which two WDs in a
    binary system coalesce due to the emission of gravitational radiation and
    ultimately explode as an SN~Ia \citep{Webbink84,Iben84}, would naturally
    account for the absence of hydrogen in the spectra.  Indeed, detection of
    {\it any} hydrogen would deal a death blow to the double-degenerate
    scenario.  Difficulties with the double-degenerate scenario are well known
    (see \citealt{Livio01} for a thorough review), however, and include both
    observational (e.g., of the $\sim 120$ double-degenerate systems known,
    none have a total mass in excess of the Chandrasekhar limit;
    \citealt{Napiwotzki04}) and theoretical (e.g., the merger process seems
    more likely to lead to collapse to a neutron star than to thermonuclear
    explosion as an SN~Ia; \citealt{Saio98}, and references therein)
    objections.  Nonetheless, theoretical loopholes remain
    \citep[e.g.,][]{Piersanti03,Yoon07}, and the observational sample of nearby
    double-degenerate systems is not complete \citep{Nelemans05}.  Since the
    double-degenerate channel is a potential ``silver-bullet'' that can explain
    both the lack of hydrogen in SNe~Ia as well as the occurrence of SNe~Ia in
    both old and young star-forming systems \citep[e.g.,][]{Branch95}, it
    requires continued theoretical and observational attention.

\section{Conclusions}
\label{sec:4}

We obtained five deep, moderate-resolution, nebular-phase spectra of two SNe~Ia
(SN~2005am and SN~2005cf) in order to search for narrow H$\alpha$ emission that
would betray the existence of material stripped from the envelope of a
mass-donating stellar companion to the exploding WD.  No such emission is
detected in either object at any epoch.  From the models of \citet{Mattila05},
we establish upper limits of $0.01\ M_\odot$ of solar abundance material in the
inner ejecta of both objects, which are the tightest constraints yet
established by such studies.  Our non-detections of H$\alpha$, coupled with the
mass-stripping results of \citet{Marietta00} and \citet{Meng07}, rule out all
hydrogen-donating companions close enough to the WD to have been experiencing
RLOF at the time of explosion for these events.  Additional theoretical work is
needed in several areas to buttress this conclusion, including most critically
verification of the transparency of the outer, more rapidly moving ejecta that
could potentially block H$\alpha$ photons from escaping from the inner region.
Bearing this caveat in mind, we propose that symbiotics are, at this time, the
most likely progenitor class that remains consistent with these data.

Definitive proof of the identity of the progenitor system(s) that gives rise to
SNe~Ia remains elusive, and it must be admitted that our conclusion, which is
based on the {\it lack} of a detection, is not as satisfying as one based {\it
on} a detection.  Should future modeling efforts prove unable to ``hide the
hydrogen'' for even widely separated binaries, then the continued viability of
the single-degenerate, hydrogen-donating progenitor will require that
H$\alpha$, no matter how weak, must ultimately be detected.

\acknowledgments

I owe a debt of gratitude to Tom Matheson for valuable assistance with the
acquisition and reduction of the data obtained with the Gemini telescopes.  I
acknowledge support from an NSF Astronomy and Astrophysics Postdoctoral
Fellowship (award AST-040147), under which part of this research was carried
out.  This research has made use of the NASA/IPAC Extragalactic Database (NED),
which is operated by the Jet Propulsion Laboratory, California Institute of
Technology, under contract with NASA.  Finally, I wish to recognize and
acknowledge the very significant cultural role and reverence that the summit of
Mauna Kea has always had within the indigenous Hawaiian community.  We are most
fortunate to have the opportunity to conduct observations from this mountain.

%\bibliography{/Users/leonard/misc/all_refs}

\clearpage

\clearpage

\input{tab1.tex}

\input{tab2.tex}

\end{document}

%% file: tab1.tex
%\documentclass[preprint]{aastex}
%\pagestyle{empty}
%\begin{document}
%Table of Spectra.
\begin{deluxetable}{llcccccrcccccc}
%\tabletypesize{\scriptsize}
\tabletypesize{\tiny}
\rotate
%\tablenum{1}
\tablewidth{650pt}
\tablecaption{Journal of Spectroscopic Observations}
\tablehead{\colhead{} & 
\colhead{} &
\colhead{} &
\colhead{HJD} &
\colhead{} &
\colhead{Range\tablenotemark{c}}  &
\colhead{Resolution\tablenotemark{d}} &
\colhead{P.A.\tablenotemark{e}} &
\colhead{Par. P.A.\tablenotemark{f}} &
\colhead{} & 
\colhead{Flux } &
\colhead{Seeing\tablenotemark{i}} &
\colhead{Slit} &
\colhead{Exposure\tablenotemark{j}} \\
\colhead{UT Date} &
\colhead{Object} &
\colhead{Day\tablenotemark{a}} &
\colhead{$-$2,400,000} &
\colhead{Telescope\tablenotemark{b}} &
\colhead{(\AA)} &
\colhead{(\AA)} &
\colhead{(deg)} &
\colhead{(deg)} &
\colhead{Air Mass\tablenotemark{g}} &
\colhead{Standard\tablenotemark{h}} &
\colhead{(arcsec)} &
\colhead{(arcsec)} &
\colhead{(s)}  }
\startdata

2005 Dec 31.52 & SN 2005am & 298.02  & 53736.02 & KI  & 5860--7160  & 3.0 & 
150 &
149--4  & 1.24--1.32 & HD84  & 1.3 & 1.0 &  7,200 \\

2006 Mar 6.59 & SN 2005cf & 267.09 & 53801.09 & GN & 5520--7628 & 3.1 & 
55 &
135--22 & 1.12--1.33 & HZ44 & 0.7 & 1.0 & 10,800 \\

2006 Mar 24.15 & SN 2005am & 380.65 & 53818.65 & GS & 5150--6920 & 3.2 & 
0 &
117--156 & 1.04--1.34 & LTT1020 & 0.7 & 1.0 & 10,410 \\

2006 Jun 1.46 & SN 2005cf & 353.96 & 53887.96 & KI & 5750--7430 & 2.4 & 
55 &
21--58 & 1.15--1.70 & BD26 & 1.1 & 1.0 & 11,100 \\

2006 Jul 1.39 & SN 2005cf & 383.89 & 53917.89 & KI & 5796--7496 & 3.4 & 
55 &
47--61 & 1.32--1.93 & BD26 & 1.0 & 1.5 & 6,600 \\

\enddata \tablecomments{All Keck observations were obtained with the
  900/5500 [number of lines mm$^{-1}$/blaze wavelength (\AA)] grating + GG495
  order blocking filter; all Gemini observations were carried out with the
  R831\_G5322 grating + OG515\_GO330 order blocking filter.  Gemini
  observations were conducted under programs GS-2006A-Q-28 (SN~2005am; PI: Leonard) and
  GN-2006A-Q-27 (SN~2005cf; PI: Leonard).}

\tablenotetext{a}{Days since estimated dates of maximum $B$ brightness. 
  SN~2005am:  2005 March $8.5 \pm 1$, HJD $2,453,438 \pm 1$ \citep{Brown05}.
  SN~2005cf:  2005 June $12.5 \pm 0.3$, HJD $2,453,534.0 \pm 0.3$
  \citep{Pastorello07}. }

\tablenotetext{b}{KI = Keck I 10 m/Low-Resolution Imaging Spectrometer +
polarimeter \citep[LRISp; ][]{Oke95}; GN(S) = Gemini North (South) 8 m/Gemini
Multi-Object Spectrograph \citep[GMOS; ][]{Hook04}.}

 \tablenotetext{c}{Wavelength range of the calibrated flux spectrum.}

\tablenotetext{d}{Approximate spectral resolution derived from night-sky lines
  near the nominal location of H$\alpha$ in the observed spectrum.}

\tablenotetext{e}{Position angle of the spectrograph slit.}

\tablenotetext{f}{Parallactic angle \citep{Filippenko82} range calculated at
  the midpoint of each separate observation for each set of observations. }

\tablenotetext{g}{Airmass range calculated at the midpoint of each separate
  observation for each set of observations.  }

\tablenotetext{h}{HD84 = HD~84937, BD26 = BD+26$^\circ$2606 \citep{Oke83}; HZ44
  = Hz 44 \citep{Massey90}; LTT1020 = LTT 1020 \citep{Baldwin84}.  Standard
  stars for observations made at the Keck Observatory were observed on the same
  night as the supernova observations; for the Gemini Observatory observations,
  the standard stars were observed on 2006 February 15 (HZ44, for the day 267
  observation of SN~2005cf) and 2005 September 5 (LTT 1020, for the day 381
  observations of SN~2005am).  The observations of LTT 1020 were made as part
  of program GS-2005B-C-4 (PI: Dan Christlein). }

\tablenotetext{i}{Average value of the full width at half maximum of the
spatial profile for each set of observations, rounded to the nearest 0\farcs1.}

\tablenotetext{j}{Combined exposure duration of all observations, in seconds.
Observations at the Keck Observatory consisted of four separate, successive
exposures, while observations at the Gemini Observatories consisted of six
separate, successive exposures.}

\label{tab:1}

\end{deluxetable}
\clearpage
%\end{document}

%% file: tab2.tex
%\documentclass[preprint]{aastex}
%\pagestyle{empty}
%\begin{document}
%Table of Spectra.
\begin{deluxetable}{lccccccc}
%\tabletypesize{\scriptsize}
\tabletypesize{\small}
%\rotate
%\tablenum{1}
%\tablewidth{350pt}
\tablecaption{Measured Values}
\tablehead{\colhead{} & 
\colhead{} &
\colhead{Scale} &
\colhead{}  &
\colhead{}  &
\colhead{$W_\lambda(3\sigma)$\tablenotemark{d}} &
\colhead{$W_\lambda (0.05 M_\odot)$\tablenotemark{e}} &
\colhead{} \\
\colhead{Object} &
\colhead{Day} &
\colhead{Factor\tablenotemark{a}} &
\colhead{$B$\tablenotemark{b}} &
\colhead{$\Delta I$\tablenotemark{c}} &
\colhead{(\AA)} &
\colhead{(\AA)} &
\colhead{${\rm M} (M_\odot)$\tablenotemark{f}} }
\startdata

SN 2005am & 298.02 & 1.16 & 1.00 & 0.090 & 2.19 & 10.76 & 0.01 \\
			  							
SN 2005cf & 267.09 & 1.45 & 0.97 & 0.015 & 0.38 & 2.11  & 0.01\\
			  							
SN 2005am & 380.65 & 0.45 & 0.94 & 0.144 & 3.73 & 21.51 & 0.01\\
			  							
SN 2005cf & 353.96 & 1.22 & 1.25 & 0.055 & 1.07 & 8.82  & 0.01\\
			  							
SN 2005cf & 383.89 & 1.02 & 0.88 & 0.164 & 4.51 & 13.10 & 0.02\\

\enddata \tablecomments{All measurements made on rest-frame spectra rebinned to
  $3$ \AA\ bin$^{-1}$.}

\tablenotetext{a}{Factor by which the original, reduced flux spectrum was
  multiplied to place it on an (approximate) absolute flux scale;
  see text for details.}

 \tablenotetext{b}{Number of 3 \AA\ bins per resolution element.}

 \tablenotetext{c}{1$\sigma$ root-mean-square fluctuation of the flux around a
normalized continuum.}

\tablenotetext{d}{Equivalent width of the strongest potential feature near
$\lambda_0 = 6563$ \AA\ that could have remained undetected in the spectrum,
derived using Equation~\ref{eqn:1}.}

\tablenotetext{e}{Expected equivalent width of an H$\alpha$ emission line
  resulting from $0.05\ M_\odot$ of solar abundance material according to the
  models of \citet{Mattila05}; see text for details.}

\tablenotetext{f}{Derived upper limit (rounded up to the nearest $0.01\
  M_\odot$) on the amount of solar abundance material that could have remained
  undetected at each spectral epoch, derived according to
  Equation~\ref{eqn:2}.}

\label{tab:2}

\end{deluxetable}
\clearpage
%\end{document}